\documentclass[a4paper,fleqn,usenatbib,useAMS]{mn2e}
\usepackage[applemac]{inputenc}
\usepackage[draft=false, colorlinks=true, allcolors=blue]{hyperref}
\usepackage{graphicx}	
\usepackage{amsmath}	
\usepackage{bm}

\newcommand{\percent}{\ensuremath{\mathrm{per\, cent}}}
\newcommand{\mh}{\ensuremath{h^{-1} \, \mathrm{M_\odot}}}

\newcommand{\kpch}{\ensuremath{h^{-1} \, \mathrm{kpc}}}
\newcommand{\cmpch}{\ensuremath{h^{-1} \, \mathrm{cMpc}}}
\newcommand{\K}{\ensuremath{\mathrm{K}}}

\newcommand{\mean}[1]{\bar{#1}}
\newcommand{\amean}[1]{\langle#1\rangle}
\newcommand{\diff}{\mathrm{d}}

\newcommand{\Om}{\ensuremath{\Omega_\mathrm{m}}}
\newcommand{\Ob}{\ensuremath{\Omega_\mathrm{b}}}
\newcommand{\Ol}{\ensuremath{\Omega_{\Lambda}}}

\newcommand{\ns}{\ensuremath{n_\mathrm{s}}}

\newcommand{\zc}{\ensuremath{\zeta_\mathrm{c}}}

\newcommand{\Rin}{\ensuremath{R_1}}
\newcommand{\Rout}{\ensuremath{R_2}}
\newcommand{\Rmax}{\ensuremath{R_\mathrm{m}}}

\defcitealias{debackere2022b}{Paper~I}

\usepackage[T1]{fontenc}
\usepackage{ae,aecompl}

\usepackage{newtxtext,newtxmath}

\title[Impact of baryons on aperture masses]{Galaxy cluster aperture masses are more robust to baryonic effects than 3D halo masses}

\author[S.N.B. Debackere et al.]{
  Stijn N.B. Debackere\thanks{Contact e-mail:
    \href{mailto:debackere@strw.leidenuniv.nl}{debackere@strw.leidenuniv.nl}
  },
  Henk Hoekstra,
  Joop Schaye
  \\
  Leiden Observatory, Leiden University, PO Box 9513, NL-2300 RA
  Leiden, The Netherlands
  \\}

\date{Last updated --; in original form --}

\pubyear{2021}

\begin{document}\label{firstpage}
\pagerange{\pageref{firstpage}--\pageref{lastpage}}
\maketitle

\begin{abstract}
  Systematic uncertainties in the mass measurement of galaxy clusters
  limit the cosmological constraining power of future surveys that
  will detect more than $10^{5}$ clusters. Previously, we argued that
  aperture masses can be inferred more accurately and precisely than
  3D masses without loss of cosmological constraining power. Here, we
  use the Baryons and Haloes of Massive Systems (BAHAMAS)
  cosmological, hydrodynamical simulations to show that aperture
  masses are also less sensitive to changes in mass caused by galaxy
  formation processes. For haloes with
  $m_\mathrm{200m,dmo} > 10^{14} \, \mh$, binned by their 3D halo
  mass, baryonic physics affects aperture masses and 3D halo masses
  similarly when measured within apertures similar to the halo virial
  radius, reaching a maximum reduction of $\approx 3 \, \percent$. For
  lower-mass haloes,
  $10^{13.5} < m_\mathrm{200m,dmo} / \mh < 10^{14}$, and aperture
  sizes $\sim 1 \, \cmpch$, representative of weak lensing
  observations, the aperture mass is consistently reduced less
  ($\lesssim 5 \, \percent$) than the 3D halo mass
  ($\lesssim 10 \, \percent$ for $m_\mathrm{200m}$). The halo mass
  reduction evolves only slightly, by up to $2$ \percent age points,
  between redshift $0.25$ and $1$ for both the aperture mass and
  $m_\mathrm{200m}$. Varying the simulated feedback strength so the
  mean simulated hot gas fraction covers the observed scatter inferred
  from X-ray observations, we find that the aperture mass is
  consistently less biased than the 3D halo mass, by up to
  $2 \, $\percent age points at
  $m_\mathrm{200m,dmo} = 10^{14} \, \mh$. Therefore, aperture mass
  calibrations provide a fruitful path to reduce the sensitivity of
  future cluster surveys to systematic uncertainties.
\end{abstract}




\section{Introduction}\label{sec:introduction}
Future large-scale surveys such as
\emph{Euclid}\footnote{\url{https://www.euclid-ec.org}} and the Rubin
Observatory Legacy Survey of Space and Time
(LSST)\footnote{\url{https://www.lsst.org/}} will study the
competition between the growth of structure from the gravitational
collapse of matter, and the accelerated expansion of the Universe due
to dark energy or modified gravity
\cite[e.g.][]{lsstsciencecollaboration2009, amendola2018}. Galaxy
clusters probe this effect particularly well because they are still
actively forming due to the hierarchical growth of structure. Hence,
the cluster abundance as a function of mass and time is sensitive to
the amount of matter and the cosmological expansion history
\citep[e.g.][]{haiman2001, allen2011}.

The statistical power of current cluster surveys is still limited by
their modest sample sizes. However, the recently released Atacama
Cosmology Telescope (ACT) cluster sample already contains $> \nolinebreak 4000$
objects \citep{hilton2021a} and for \emph{Euclid} and the LSST sample
sizes of $>10^5$ objects are expected \citep[e.g.][]{tyson2003,
  sartoris2016}, ushering in the era of cluster surveys that will be
limited by systematic uncertainties \citep{kohlinger2015}.

Currently, building a cluster sample for a cosmology analysis requires
three steps. First, clusters need to be detected in the data by
identifying large matter overdensities either through the clustering
of galaxies in space and redshift in optical images or through peaks
in the X-ray emission, the Sunyaev-Zel'dovich effect or the weak
lensing shear. Second, measures of cluster masses are calibrated,
usually, by measuring the mass--observable relation that links the
survey detection observable to the cluster mass derived from weak
lensing observations. Third, by modelling the cluster selection
through the survey observable, the measured abundance can be compared
to predictions based on the theoretical, cosmology-dependent halo mass
function to constrain the cosmological parameters of the Universe.

The exponential sensitivity of the cluster abundance to the cluster
mass means that the accuracy of the cluster mass calibration limits
the cosmological constraining power of cluster surveys. In
\citet{debackere2022b}, hereafter \citetalias{debackere2022b}, we
argued that aperture mass calibrations can greatly reduce the
systematic uncertainty of cluster cosmology analyses. Aperture masses
can be measured directly from weak lensing observations and in
simulations, which avoids the deprojection of the observations and,
hence, bypasses the additional biases and uncertainties introduced by
the assumed spherically symmetric density profile in the deprojection.
Moreover, aperture masses can be measured within fixed angular or
physical apertures, with no need to derive an overdensity radius that
depends on the assumed density profile. The aperture mass measurement
uncertainty depends solely on the number of background galaxies used
to sample the shear field and is $\approx 2-3$ times smaller than the
uncertainty in the inferred 3D halo masses. We also showed that,
compared to the 3D halo mass function, the aperture mass function is
more sensitive to changes in $\Om$ and $w_a$, similarly sensitive to
changes in $\sigma_8$, $w_0$, and $n_\mathrm{s}$, and slightly less
sensitive to changes in $h$. Hence, the aperture mass function can
also constrain the cosmological evolution of the Universe.

Since the projected mass within an aperture is actually derived from
the weak lensing signal of galaxies outside of the aperture, the
measured aperture mass is also less sensitive to sources of systematic
error near the cluster centre such as miscentring and contamination of
the lensing signal due to cluster galaxies
\citep[e.g.][]{mandelbaum2010a, hoekstra2012}. Moreover, aperture
masses can be measured unambiguously, even for triaxial and disturbed
systems. Hence, the aperture mass measurement is relatively robust to
different sources of systematic uncertainty.

One source of systematic uncertainty that will become important for
future surveys, is the impact of baryonic physics on the inferred
cluster mass compared with a universe containing only dark matter. All
currently available theoretical predictions of the halo abundance rely
on suites of large-volume dark matter-only (DMO) simulations
\citep[e.g.][]{tinker2008, nishimichi2018, mcclintock2019a,
  bocquet2020}. However, we have known for a long time that baryonic
processes related to galaxy formation can significantly modify cluster
masses \citep[e.g.][]{rudd2008, stanek2009, cui2012, martizzi2014,
  velliscig2014, bocquet2016}. To ensure realistic cluster gas
fractions---and to prevent overcooling---simulations with subgrid
models for radiative cooling and star formation also need to include
the feedback from Active Galactic Nuclei \citep[AGN,
e.g.][]{mccarthy2010}. For clusters with masses between
$10^{14} \lesssim m_\mathrm{200m}/(\mh) \lesssim 10^{14.5}$, where
$m_\mathrm{200m}$ is the mass within a radius enclosing an average
overdensity \footnote{In \citet{debackere2021}, we wrongly defined
  this spherical overdensity as
  $\amean{\rho}=200\Om \rho_\mathrm{crit}(z)$. However, the actual
  mass calculations used in that paper \emph{did} use the correct
  overdensity definition $\amean{\rho}=200 \rho_\mathrm{m}(z)$, as can
  be seen from the
  \href{https://github.com/StijnDebackere/lensing_haloes/blob/afe4f8699003c27dbf1298c343dbe473d47fd165/lensing_haloes/results.py\#L312}{publicly
    available code.}} of
$\amean{\rho} = 200 \rho_\mathrm{m}(z) = 200 \Om
\rho_\mathrm{crit}(z=0)(1+z)^3$, AGN feedback reduces the total halo
mass by $1-5 \, \percent$ with a larger impact for lower halo masses,
compared to the same halo in a universe comprising only dark matter
\citep[e.g.][]{velliscig2014, bocquet2016}. This mass reduction will
become an important systematic uncertainty due to the increased
statistical power of future surveys. Since the aperture mass measures
the projected mass, it should be less sensitive to baryonic processes
that dominate the cluster density profile on small scales \citep[see
e.g.][]{henson2017, lee2018, debackere2021}.

To include the effect of baryons in a traditional cluster cosmology
analysis that relies on DMO simulations to predict the cosmology
dependence of the halo mass function, we need to include a baryonic
correction in the theoretical halo mass to infer unbiased cosmological
parameters \citep[e.g.][]{balaguera-antolinez2013, debackere2021}.
Thus, to write down the full forward model of the observed cluster
number counts, we need a calibrated mass--observable relation,
$P(\mathcal{O}|\mathcal{M})$, a theoretical prediction of the
cosmology-dependent halo abundance, $n(\mathcal{M}, \mathbf{\Omega})$,
where $\mathbf{\Omega}$ indicates the cosmological parameters, and a
conversion between the observed cluster mass,
$\mathcal{M}_\mathrm{obs}$, and the theoretical halo mass,
$\mathcal{M}$. This conversion includes the baryonic mass correction
if the theoretical halo abundance was predicted using DMO simulations.
We can then write the number counts,
$N(\mathcal{O}_i, z_j|\mathbf{\Omega})$, within the observable bin,
$\mathcal{O}_i$, and redshift bin, $z_j$, for the assumed cosmology,
$\mathbf{\Omega}$, as
\begin{equation}
  \label{eq:N_obs}
  N(\mathcal{O}_i, z_j| \mathbf{\Omega}) = \Omega_\mathrm{sky} 
  \begin{aligned}[t]
    & \int\limits_{\mathcal{O}_i}^{\mathcal{O}_{i+1}} \diff
      \mathcal{O} \int\limits_{z_j}^{z_{j+1}} \diff z \int
      \diff \mathcal{M}_\mathrm{dmo} \mathcal{M}_\mathrm{hydro} \, \diff \mathcal{M}_\mathrm{obs} \\
    & \times P(\mathcal{O}|\mathcal{M}_\mathrm{obs}, z)
      P(\mathcal{M}_\mathrm{obs}|\mathcal{M}_\mathrm{hydro}, z)\\
    & \times P(\mathcal{M}_\mathrm{hydro}|\mathcal{M}_\mathrm{dmo}, z)
      n_\Omega(\mathcal{M}_\mathrm{dmo}, z|\mathbf{\Omega}) \, .
  \end{aligned}
\end{equation}
Here, the observable and the redshift are integrated over their
respective bins and the different halo masses from $0$ to $\infty$.
The theoretical halo mass function, $n_\Omega$, is calculated per unit
survey area and redshift interval. We have introduced a
redshift-dependent conversion between the theoretical halo mass from
DMO simulations, $\mathcal{M}_\mathrm{dmo}$, and the mass of the same
halo in a universe containing baryons, $\mathcal{M}_\mathrm{hydro}$.
Moreover, we explicitly differentiate between
$\mathcal{M}_\mathrm{obs}$, the cluster mass measured observationally,
and $\mathcal{M}_\mathrm{hydro}$, the total halo mass measured in the
hydrodynamical simulation, since $\mathcal{M}_\mathrm{obs}$ is a noisy
measurement of $\mathcal{M}_\mathrm{hydro}$ due to observational
systematic uncertainties. We will assume perfect knowledge of the
selection function to simplify the analysis, but we refer the
interested reader to Section 5.1 of \citetalias{debackere2022b} for a
discussion about how the selection function will impact a cosmological
analysis relying on aperture mass calibrations.

One straightforward way to eliminate the systematic uncertainty in
converting halo masses from the DMO to the hydrodynamical simulation,
is to predict the cosmology dependence of the halo abundance directly
from large-volume hydrodynamical simulations for a grid of
cosmological parameters, i.e. to predict
$n_\Omega(\mathcal{M}_\mathrm{hydro}, z|\mathbf{\Omega})$. However,
due to the computational expense and the uncertain astrophysics, such
an effort has so far not been undertaken.

The relation between the measured observable and halo mass,
$P(\mathcal{O}|\mathcal{M}_\mathrm{obs})$ can be measured
observationally, with the caveat that the inferred 3D halo mass
depends on the density profile assumed in the deprojection, thus
introducing a significant modelling uncertainty. The measurement
uncertainty, $P(\mathcal{M}_\mathrm{obs}|\mathcal{M}_\mathrm{hydro})$,
can be calibrated using simulations for both the aperture mass and the
3D halo mass. In \citetalias{debackere2022b}, we emphasized that the
measurement uncertainty in the aperture mass,
$P(\Delta M_\mathrm{obs}|\Delta M)$, depends only on the number
density of background galaxies used to reconstruct the weak lensing
shear. For 3D halo masses, on the other hand, the assumption of a
density profile to deproject the observations and to infer the mass
within a fixed overdensity radius, introduces a model-dependent bias
in the inferred halo mass due to the mismatch between the spherically
symmetric density profile and the true, triaxial halo, including
substructure and correlated structure. Uncorrelated structure along
the line-of-sight introduces an additional uncertainty
\citep[e.g.][]{hoekstra2001}.

We use the Baryons and Haloes of Massive Systems (BAHAMAS) suite of
large-volume cosmological, hydrodynamical simulations
\citep{mccarthy2017a, mccarthy2018} to study the effect of feedback
processes related to galaxy formation on halo aperture masses, that
is, the $P(\mathcal{M}_\mathrm{hydro}|\mathcal{M}_\mathrm{dmo})$ term
in Eq.~\eqref{eq:N_obs}. The BAHAMAS simulations have been calibrated
to reproduce the observed galaxy stellar mass function and the cluster
hot gas fractions derived from X-ray observations, and they reproduce
a wide range of observed properties of massive systems, enabling
realistic cosmology forecasts that include the effect of baryons. We
quantify the change in the aperture mass in BAHAMAS, and examine how
it depends on the strength of the implemented feedback. We compare our
results to the baryonic correction to the 3D DMO halo mass.

This paper is structured as follows: in Section~\ref{sec:sims}, we
introduce the BAHAMAS simulations, describe the aperture mass
measurement, and we discuss the matching between haloes in the
hydrodynamical and DMO simulations. In
Section~\ref{sec:m_ap_behaviour}, we show the relation between the
aperture mass and the 3D halo mass. In
Section~\ref{sec:mass_correction}, we compare the mean change in the
halo mass when including baryons and its scatter for both aperture
masses and 3D halo masses, we study its redshift evolution and
sensitivity to different baryonic physics models that bracket the
observationally allowed range of cluster gas fractions derived from
X-ray observations. We conclude in Section~\ref{sec:conclusions}.

\section{Simulations}\label{sec:sims}
We measured the projected aperture masses of group and cluster-sized
haloes from the BAHAMAS suite of cosmological hydrodynamical
simulations \citep{mccarthy2017a}. This suite of simulations is
well-suited for our aims for several reasons. First, due to the
$(400 \cmpch)^3$ volume, we obtain a sufficiently large sample of
massive haloes with $m_\mathrm{200m} > 10^{13.5} \mh$. Second, the
subgrid model parameters for the feedback from supernovae and AGN of
the fiducial simulation have been calibrated to reproduce the
present-day galaxy stellar mass function (GSMF), and, crucially for
our work, the hot gas mass fractions of groups and clusters of
galaxies. Moreover, variations of both the cosmological model and of
the non-resolved, subgrid physics model parameters are available.

\subsection{Simulation set}\label{sec:sim_list}
The BAHAMAS model remains unchanged from its predecessors OWLS
\citep{schaye2010} and cosmo-OWLS \citep{lebrun2014}, except for the
values of the subgrid model parameters, which were chosen to reproduce
the observed large-scale mass distribution of the Universe. Hence, we
refer the interested reader to \citet{schaye2010} for a detailed
description of the different subgrid physics models.

The BAHAMAS suite consists of simulations run with a modified version
of the Lagrangian TreePM-SPH code \textsc{Gadget-3}
\citep[unpublished--for \textsc{Gadget-2}, see][]{springel2005a} in
boxes with a periodic side length of $400 \cmpch$ with initial
conditions matching the maximum-likelihood cosmological parameter
values from the WMAP9 data \citep{hinshaw2013}, i.e.
$\{\Om, \Ob, \Ol, \sigma_8, \ns, h\} = \{0.2793, 0.0463, 0.7207,
0.821, 0.972, 0.700\}$. The initial linear power spectrum is generated
at $z=127$ using CAMB\footnote{https://camb.info/} \citep{lewis2000}
and converted into particle positions using
S-GenIC\footnote{https://github.com/sbird/S-GenIC}, a modified version
of NGenIC\footnote{https://www.h-its.org/2014/11/05/ngenic-code/},
that includes second-order Lagrangian perturbation theory and supports
massive neutrinos. The hydrodynamical and their corresponding dark
matter-only (DMO) simulations contain $(2 \times 1024)^3$ and $1024^3$
particles, respectively. This results in dark matter and (initial)
baryonic particle masses of $\approx 3.85 \times 10^9 \mh$ and
$\approx 7.66 \times 10^8 \mh$, respectively, for the WMAP9 cosmology
(the dark matter particle mass in the DMO simulations is
$\approx 4.62 \times 10^9 \mh$). The gravitational softening length is
set to $4 \kpch$ in physical (comoving) coordinates for
$z \leq (>) 3$.

Haloes are identified using the Friends-of-Friends (FoF) algorithm
with a linking length of $0.2$ and their spherical overdensity masses
are calculated from all the particles within the FoF halo, including
particles that are not gravitationally bound, centred on the minimum
of the gravitational potential using \textsc{Subfind}
\citep{springel2001a}. The so-called subgrid models for non-resolved
physical processes were taken from the preceding OWLS and cosmo-OWLS
projects \citep[][respectively]{schaye2010, lebrun2014}. These models
include recipes for the radiative heating and cooling of the 11
dominant elements tracked in the simulations (H, He, C, N, O, Ne, Mg,
Si, S, Ca, Fe), by interpolating the tabulated \textsc{Cloudy}
\citep{ferland1998} rates of \citet{wiersma2009a} as a function of
density, temperature and redshift. Star formation follows the
implementation of \citet{schaye2008}, fixing the unresolved cold
interstellar medium (ISM) gas to an effective equation of state and a
pressure-dependent star formation efficiency in order to reproduce the
observed Kennicut--Schmidt star formation law. Stellar evolution and
the chemical enrichment of gas due to both type Ia and type II
supernovae, stellar winds, and asymptotic giant branch (AGB) stars are
implemented following \citet{wiersma2009}. Supernova feedback is
implemented kinetically, following \citet{dallavecchia2008}. Finally,
black hole seeding in low-mass galaxies, black hole growth through
mergers and gas accretion, and the feedback from active galactic
nuclei are modelled following \citet{booth2009}.

In Table~\ref{tab:sim_list}, we list the specific simulations of the
BAHAMAS suite that we use in this work. We list the DMO simulation and
the hydrodynamical simulations with identical initial conditions and
possible variations in the subgrid model assumptions. We will
investigate the impact of variations in the strength of the AGN
feedback by increasing (decreasing) the heating temperature
$\Delta T_\mathrm{heat}$ by $0.2 \, \mathrm{dex}$ relative to the
calibrated, fiducial value of
$\Delta T_\mathrm{heat} = 10^{7.8} \, \mathrm{K}$. This results in
lower (higher) hot gas mass fractions in groups and clusters of
galaxies \citep[see][]{mccarthy2017a}.

\begin{table}
  \caption{A list of all the simulations (dark matter-only and the
    matching hydrodynamical runs) for which we computed the halo
    aperture masses. BAHAMAS simulation names follow the convention
    TYPE\_nuN\_ZZZ, with N the sum of the neutrino masses in
    $\mathrm{eV}$ and ZZZ the base cosmological model. All simulations
    have periodic side lengths of $400 \cmpch$ and $1024^3$ dark
    matter particles (with the same number of baryonic particles in
    the hydrodynamical case). }
  \centering
  \begin{tabular}{l c c}
    \hline
    simulation & redshifts & variation \\
    \hline
    DMONLY\_nu0\_WMAP9 & 0.25, 0.5, 1 & ---  \\
    AGN\_TUNED\_nu0\_WMAP9 & 0.25, 0.5, 1 & $\Delta T_\mathrm{heat} = 10^{7.8} \, \K$ \\
    AGN\_7p6\_nu0\_WMAP9 & 0.25, 0.5, 1 & $\Delta T_\mathrm{heat} = 10^{7.6} \, \K$ \\
    AGN\_8p0\_nu0\_WMAP9 & 0.25, 0.5, 1 & $\Delta T_\mathrm{heat} = 10^{8.0} \, \K$ \\
  \end{tabular}
  \label{tab:sim_list}
\end{table}

\subsection{Aperture mass measurement}\label{sec:aperture_calc}
We follow the literature and refer to the excess projected mass within
an aperture of size $\Rin$, defined as
\begin{align}
  \label{eq:delta_m}
  \Delta M(< \Rin| \Rout, \Rmax) & = \pi \Rin^2(\mean{\Sigma}(< \Rin) - \mean{\Sigma}(\Rout < R < \Rmax)) \\
  \nonumber
                                 & = M(<\Rin) - M_\mathrm{bg}(<\Rin) \, ,
\end{align}
as the aperture mass \citep[e.g.][]{bartelmann2001a}. The background
surface mass density within $\Rin$ is inferred from the annulus
between $\Rout$ and $\Rmax$. We have introduced the mean surface mass
density
\begin{equation}
  \label{eq:sigma_mean}
  \mean{\Sigma}(\Rout < R < \Rmax) = \frac{2}{(\Rmax^2 - \Rout^2)} \int_{\Rout < R < \Rmax} \diff R\, R \Sigma(R) \, .
\end{equation}
In Appendix A of \citetalias{debackere2022b}, we showed that this
definition of the aperture mass matches the $\zc$-statistic
\citep{clowe1998}, which measures the enclosed excess surface mass
density within $\Rin$ from the observed weak lensing galaxy shears
between $\Rin$ and $\Rmax$.

For our analysis, we generated projected surface mass density maps
from the full simulation volume for each simulation in
Table~\ref{tab:sim_list}. First, we projected all the particles along
the three principal axes of the simulation box. Then, we binned the
projected particles into a pixel grid of
$0.05 \cmpch \times 0.05\cmpch$ resolution and obtained the surface
mass density $\Sigma(i,j)$ for pixel $(i,j)$ by summing the masses for
all particles with coordinates $(x,y)$ belonging to the pixel $(i, j)$
and dividing by the pixel area. From the surface mass density, we
calculated the aperture mass using Eq.~\eqref{eq:delta_m} with
$\Rin = [0.5, 1.0, 1.5] \, \cmpch$ and
$(\Rout, \Rmax) = (2, 3) \, \cmpch$, centred on the potential minimum,
for all haloes with $m_\mathrm{200m,dmo} > 10^{13} \, \mh$. The chosen
apertures are representative of weak lensing observations
\citep[e.g.][]{hoekstra2012, applegate2014}.

\subsection{Matching haloes to their DMO counterparts}\label{sec:hydro_dmo_linking}
To quantify the influence of baryons on the halo aperture masses, we
compare the aperture masses from haloes in the hydrodynamical
simulations to those of their counterparts in a universe including
only dark matter particles. (Technically, baryons are included in the
transfer function used to calculate the initial conditions.) Since all
BAHAMAS simulations with the same cosmological model have identical
initial conditions and consistent, unique dark matter particle
identification numbers, we can link haloes between the DMO and
hydrodynamical simulations. We follow the linking method of
\citet{velliscig2014}. Briefly, we identify each halo in the reference
simulation to the halo in the matching simulation that contains at
least half of its $N_\mathrm{mb}=50$ most-bound particles. Only if the
same haloes are also linked when swapping the reference and the
matching simulation in this procedure, do we consider them genuine
counterparts. Haloes with $m_\mathrm{200m,dmo} > 10^{13} \, \mh$ are
matched with a success rate higher than $98\, \percent$ in all
simulations.

One important caveat, which is especially important for aperture mass
measurements, is that the dynamical history can differ between matched
haloes in the hydrodynamical and the DMO simulation. Star formation
and feedback processes modify the matter distribution, even though the
distribution of haloes statistically remains the same on scales larger
than the halo virial radius \citep[e.g.][]{vandaalen2014}. The median
3D offset, $\Delta r$, between the $3974$ matched haloes with
$m_\mathrm{200m,hydro} > 10^{13.5} \, \mh$ is $\approx 0.1 \cmpch$,
with $17$ $(151)$ haloes having $\Delta r > 1 \, (0.2) \, \cmpch$.
Upon visual inspection, the majority of the systems with large offsets
are mergers where the haloes are identified as different components
between the hydrodynamical and the DMO simulation, resulting in
significantly different aperture mass measurements.

We exclude all haloes with 3D offsets $\Delta r > 0.2 \, \cmpch$ from
our sample. Given the minimum aperture size of $\Rin = 0.5 \, \cmpch$,
this cutoff ensures that the aperture mass measurements are not
significantly affected by possibly misidentified haloes. The main
effect of this selection criterion is to slightly reduce the scatter
in the baryonic mass correction at the high 3D halo mass end, as some
of the most massive haloes in BAHAMAS are merging and do not visually
match between the hydrodynamical and the DMO simulation.

\section{The relation between aperture mass and 3D halo mass}\label{sec:m_ap_behaviour}
\begin{figure}
  \centering
  \includegraphics[width=\columnwidth]{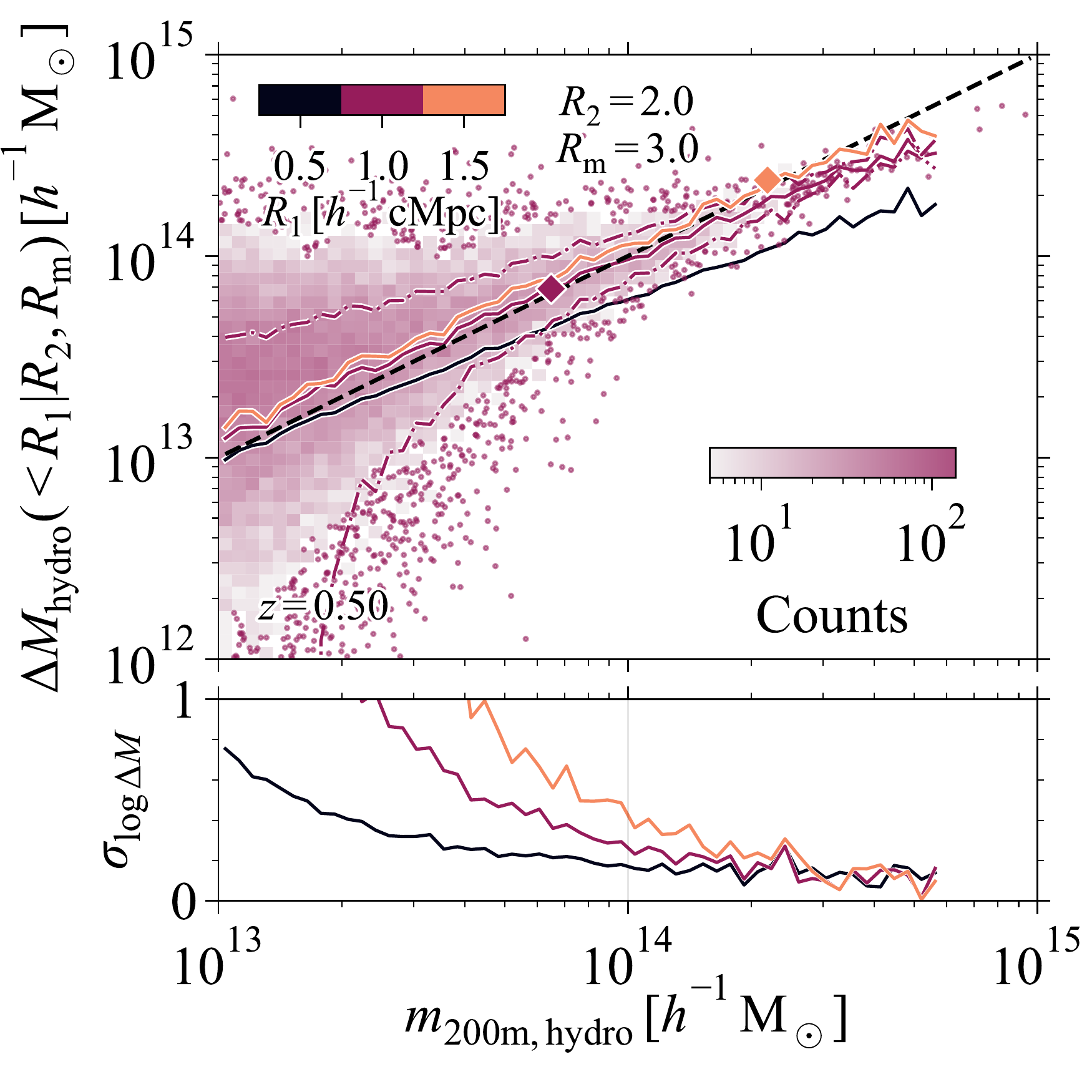}
  \caption{\emph{Top panel:} The distribution of aperture masses
    within $1\cmpch$,
    $\Delta M_\mathrm{hydro}(<1 \, \cmpch|\Rout = 2 \, \cmpch, \Rmax =
    3 \, \cmpch)$, as a function of the 3D halo mass,
    $m_\mathrm{200m,hydro}$, at $z=0.5$. The median relations for
    different aperture sizes are shown as coloured lines, with the
    dash-dotted lines indicating the 16th to 84th percentile scatter
    for $\Rin=1\,\cmpch$. The one-to-one relation is indicated by the
    black, dashed line. The coloured diamonds indicate the halo mass
    where $r_\mathrm{200m} = \Rin$. There is a large scatter in
    aperture mass at fixed, low $m_\mathrm{200m,hydro}$ due to the
    variation in the structure along the line-of-sight to different
    haloes. The aperture mass tends to be slightly higher (lower) than
    the 3D halo mass when $\Rin > \, (<) r_\mathrm{200m,hydro}$.
    \emph{Bottom panel:} The logarithmic scatter in the aperture mass
    at fixed $m_\mathrm{200m,hydro}$, measured as half the difference
    between the 84th and 16th percentiles. The scatter increases for
    lower 3D halo masses since matter outside the halo dominates the
    aperture mass. For
    $m_\mathrm{200m,hydro} \lesssim 10^{13.5} \, \mh$, the scatter
    increases significantly since $> 5\,\percent$ of the haloes is
    surrounded by more massive structures resulting in negative
    aperture masses.}\label{fig:m_ap_vs_m200m}
\end{figure}
In Fig.~\ref{fig:m_ap_vs_m200m}, we show the full distribution of
projected masses within apertures of size $\Rin = 1 \, \cmpch$ as a
function of the 3D halo mass, $m_\mathrm{200m,hydro}$, for all haloes
at $z=0.5$ in the AGN\_TUNED\_nu0\_WMAP9 simulation. The median
aperture mass at fixed 3D halo mass is shown with different coloured
lines for different apertures $\Rin$. We indicate the halo mass for
which $r_\mathrm{200m}=\Rin$ with coloured diamonds. Within a fixed
aperture, aperture masses are slightly higher than the 3D halo mass
when $\Rin > r_\mathrm{200m}$ and lower when $\Rin < r_\mathrm{200m}$
as the halo mass represents a smaller or larger fraction of the total
aperture mass, respectively. Larger apertures result in larger masses.
For masses $m_\mathrm{200m,hydro} \lesssim 10^{14} \, \mh$ a small but
non-negligible fraction of the haloes will be surrounded by more
massive structures along the line-of-sight, resulting in negative
aperture masses. The fraction of haloes with negative aperture mass
within $\Rin = 1 \, \cmpch$ increases from $0 \, \percent$ for
$m_\mathrm{200m,hydro} = 10^{14} \, \mh$ to $\approx 5 \, \percent$
for $m_\mathrm{200m,hydro} = 10^{13.5} \, \mh$. The fraction of
negative aperture masses increases with increasing aperture size.

From the bottom panel of Fig.~\ref{fig:m_ap_vs_m200m}, we can see that
the scatter in the aperture mass at $\Rin=1\,\cmpch$ at fixed 3D halo
mass, calculated as half the difference between the 84th and the 16th
percentiles of the aperture mass at fixed 3D mass, increases from
$\sigma_{\log \Delta M} \approx 0.15$ for
$m_\mathrm{200m,hydro} = 10^{14.5} \, \mh$ to $\approx 0.3$ for
$10^{14} \, \mh$. Smaller apertures result in smaller scatter. We
discussed in \citetalias{debackere2022b} that the increase in the
scatter of the aperture mass measured within a fixed aperture with
decreasing 3D halo mass is caused by the large variation in the
correlated structure surrounding the halo, which contributes more
significantly to the total aperture mass for lower-mass haloes.

\section{Aperture mass correction due to baryonic effects}\label{sec:mass_correction}
We compare the aperture mass for matched haloes between the
hydrodynamical and the DMO simulation to study the change in mass due
to the inclusion of baryons and their associated galaxy formation
processes. As shown by Eq.~\eqref{eq:N_obs}, the change in the cluster
mass due to baryons can be included in the forward model of the
cluster abundance when the cosmology dependence is predicted using DMO
simulations. The relevant term in Eq.~\eqref{eq:N_obs},
$P(\mathcal{M}_\mathrm{hydro}|\mathcal{M}_\mathrm{dmo}, z)$,
complicates the analysis since baryons introduce a mass and possibly
redshift-dependent bias between $\mathcal{M}_\mathrm{dmo}$ and
$\mathcal{M}_\mathrm{hydro}$. Moreover, the uncertainty in
$\mathcal{M}_\mathrm{hydro}$ at fixed $\mathcal{M}_\mathrm{dmo}$ needs
to be accounted for correctly in order to convert the theoretical halo
mass, $\mathcal{M}_\mathrm{dmo}$, to the cluster observable,
$\mathcal{O}$.

\subsection{Binned by 3D halo mass}\label{sec:mass_correction_3d}
\begin{figure}
  \centering
  \includegraphics[width=\columnwidth]{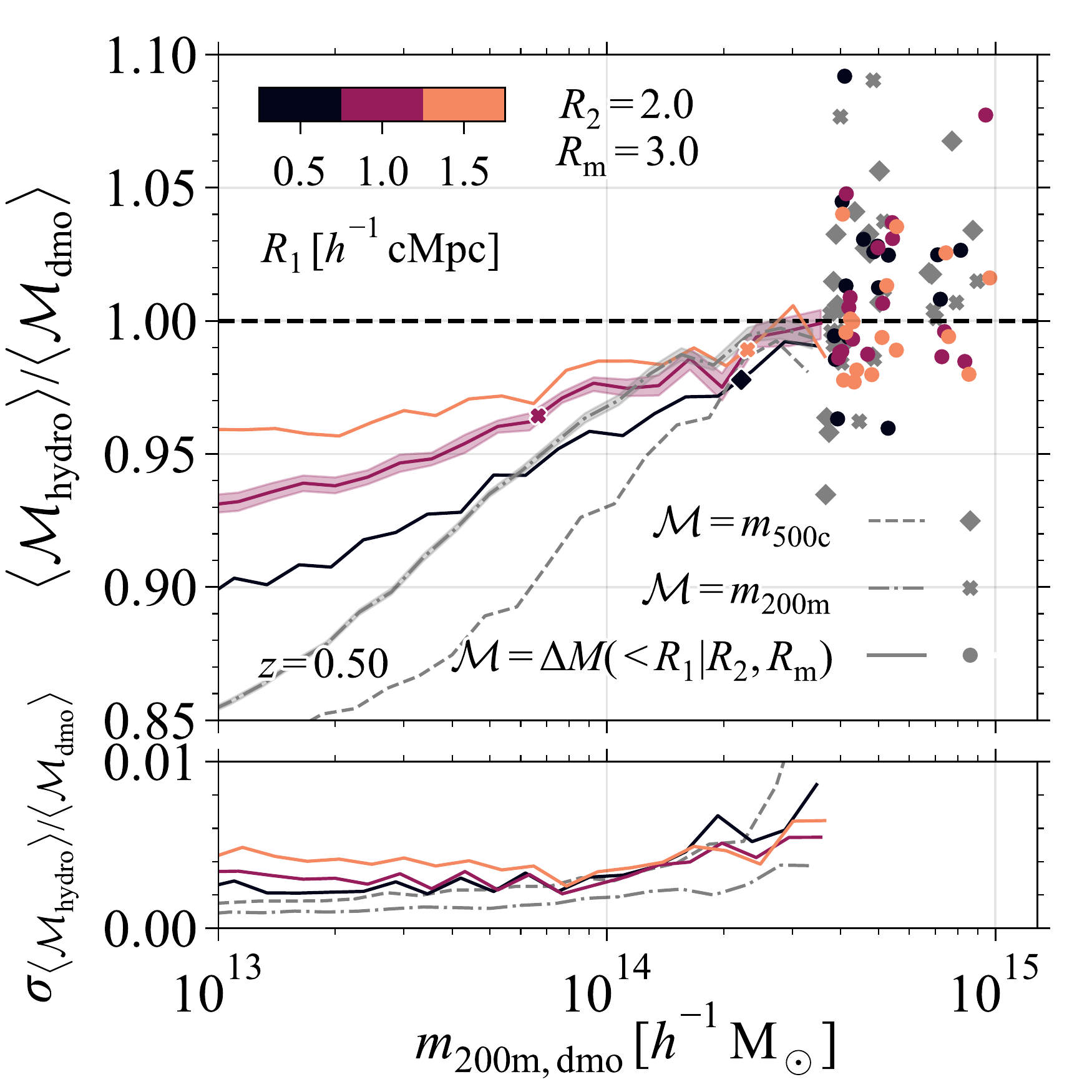}
  \caption{\emph{Top panel:} The mean aperture mass of haloes in the
    reference hydrodynamical simulation relative to the mean mass of
    the matched dark matter-only simulation counterparts at $z=0.5$,
    stacked in bins of the 3D dark matter-only halo mass. Different
    coloured lines show different aperture sizes
    $\Rin=[0.5, 1, 1.5] \, \cmpch$, with the background contribution
    calculated between $\Rout=2 \, \cmpch$ and $\Rmax=3 \, \cmpch$,
    representative for weak lensing observations. Dash-dotted (dashed)
    gray lines show the change in the 3D halo mass
    $\amean{m_\mathrm{200m,hydro}}/\amean{m_\mathrm{200m,dmo}}$
    ($\amean{m_\mathrm{500c,hydro}}/\amean{m_\mathrm{500c,dmo}}$).
    Shaded regions show the bootstrapped error on the ratio of the
    mean masses for $m_\mathrm{200m}$ and
    $\Delta M(<1 \, \cmpch|\Rout, \Rmax)$. For bins with fewer than
    $10$ haloes, individual measurements are shown with coloured
    points for the aperture mass, and gray crosses (diamonds) for the
    3D halo mass, $m_\mathrm{200m}$ ($m_\mathrm{500c}$). Coloured
    crosses (diamonds) show the 3D halo mass for which
    $r_\mathrm{200m,dmo} = \Rin$ ($r_\mathrm{500c,dmo} = \Rin$). For
    $m_\mathrm{200m,dmo} \gtrsim 10^{14} \, \mh$, the mass change is
    $< 5\, \percent$ for all mass measures. For halo masses
    $\lesssim 10^{14} \, \mh$, the aperture mass is consistently less
    biased than the 3D halo mass. \emph{Bottom panel:} The
    $1\, \sigma$ bootstrapped uncertainty in the mean mass change is
    $< 1 \, \percent$ for all mass measurements. For high-mass haloes
    the decrease in the number of haloes causes the larger
    uncertainty.}\label{fig:m_ap_vs_m200m_correction}
\end{figure}
In a cluster cosmology analysis, baryonic effects enter as a
correction in the theoretical, DMO halo mass, given by
$P(\mathcal{M}_\mathrm{hydro}|\mathcal{M}_\mathrm{dmo}, z)$ in
Eq.~\eqref{eq:N_obs}. We compare the correction in the aperture mass
and the 3D halo mass for the same halo sample by binning haloes
according to their 3D DMO halo mass. We note that an analysis that
uses the aperture mass function to model the cosmology-dependence of
the cluster sample, needs to bin the halo sample by the aperture mass
to model the correction, as we do in
Section~\ref{sec:mass_correction_aperture}.

To calculate the mean mass correction of the halo sample binned by the
3D DMO halo mass, we compute the ratio between the mean stacked halo
masses of the matched haloes in the hydrodynamical and DMO
simulations, i.e.
$\amean{\mathcal{M}_\mathrm{hydro}} /
\amean{\mathcal{M}_\mathrm{dmo}}$. Especially for the aperture mass,
it is important to use the ratio of the mean masses instead of the
mean of the mass ratios of individual haloes, i.e.
$\amean{\mathcal{M}_\mathrm{hydro} / \mathcal{M}_\mathrm{dmo}}$,
because of the large scatter in the aperture mass at fixed 3D halo
mass (see Fig.~\ref{fig:m_ap_vs_m200m}). Low-aperture mass haloes
contribute a disproportionately large uncertainty to
$\amean{\mathcal{M}_\mathrm{hydro} / \mathcal{M}_\mathrm{dmo}}$ since
a small difference in the projected mass, due to the different halo
dynamical history in the hydrodynamical and the DMO simulations,
causes large fluctuations in the individual mass ratios. These
low-aperture mass haloes do not contribute significantly to the mean
mass in the halo stack, minimizing their impact on
$\amean{\mathcal{M}_\mathrm{hydro}} /
\amean{\mathcal{M}_\mathrm{dmo}}$.

In the top panel of Fig.~\ref{fig:m_ap_vs_m200m_correction}, we show
the mean aperture mass in the hydrodynamical simulation relative to
the DMO simulation in a stack of haloes binned by their 3D DMO halo
mass, $m_\mathrm{200m,dmo}$. The different coloured lines show the
binned aperture mass changes for different aperture sizes, and the
gray, dash-dotted and dashed lines for $m_\mathrm{200m}$ and
$m_\mathrm{500c}$, respectively. Since the halo baryon fraction
increases with the halocentric distance, the baryonic correction
decreases for larger apertures and radii enclosing a smaller
overdensity. Since more massive haloes are able to retain a larger
fraction of the cosmic baryons, the mass change also decreases with
increasing 3D halo mass. For the most massive haloes,
$m_\mathrm{200m,dmo} \gtrsim 10^{14.5} \, (10^{14}) \, \mh$, the mass
reduction is $\lesssim 1 \, (5) \, \percent$ for all halo mass
measures. For lower-mass haloes, the aperture mass is consistently
less biased than the 3D halo mass.

The halo mass change due to the inclusion of baryons is caused by the
heating of the intracluster gas by AGN feedback and galactic winds,
transporting baryons to the halo outskirts and reducing the inner halo
baryon fraction \citep[e.g.][]{velliscig2014}. For this reason, mass
measurements that include more of the outer halo density profile will
differ less from the DMO halo mass. We can see from
Fig.~\ref{fig:m_ap_vs_m200m_correction} that aperture masses are less
sensitive to the impact of baryons than 3D halo masses as masses
measured within $\Rin = 0.5 \, \cmpch$ are less biased than
$m_\mathrm{200m}$ even when $r_\mathrm{200m,dmo} > 0.5 \, \cmpch$
because aperture masses probe scales larger than $\Rin$ along the
projection axis.

In the bottom panel of Fig.~\ref{fig:m_ap_vs_m200m_correction}, we
show the bootstrapped uncertainty in the ratio between the mean halo
masses measured in the hydrodynamical and DMO simulations. We obtain
the bootstrapped distribution of the mean halo mass in each bin of
$m_\mathrm{200m,dmo}$ by resampling the haloes 500 times with
replacement. Then, we calculate the uncertainty as half the difference
between the 84th and the 16th percentiles. We also show the
uncertainty as the shaded region in the top panel of
Fig.~\ref{fig:m_ap_vs_m200m_correction} for the cases
$\Rin=1 \, \cmpch$ and $m_\mathrm{200m}$. The uncertainty in the mean
mass change is smallest for $m_\mathrm{200m}$, being between $1.5$ to
$2$ times smaller than the uncertainty in $m_\mathrm{500c}$ and the
aperture mass measured within $\Rin \leq 1 \, \cmpch$. The increase in
the uncertainty for higher-mass haloes is due to the limited sample
size of the BAHAMAS simulation. The mass change for all mass
measurements can be determined with sub\percent\ accuracy with the
BAHAMAS cluster sample of $\approx 30 \, (3800)$ haloes with
$m_\mathrm{200m,hydro} > 10^{14.5} \, (10^{13.5})\, \mh$.

\subsection{Binned by aperture mass}\label{sec:mass_correction_aperture}
\begin{figure}
  \centering
  \includegraphics[width=\columnwidth]{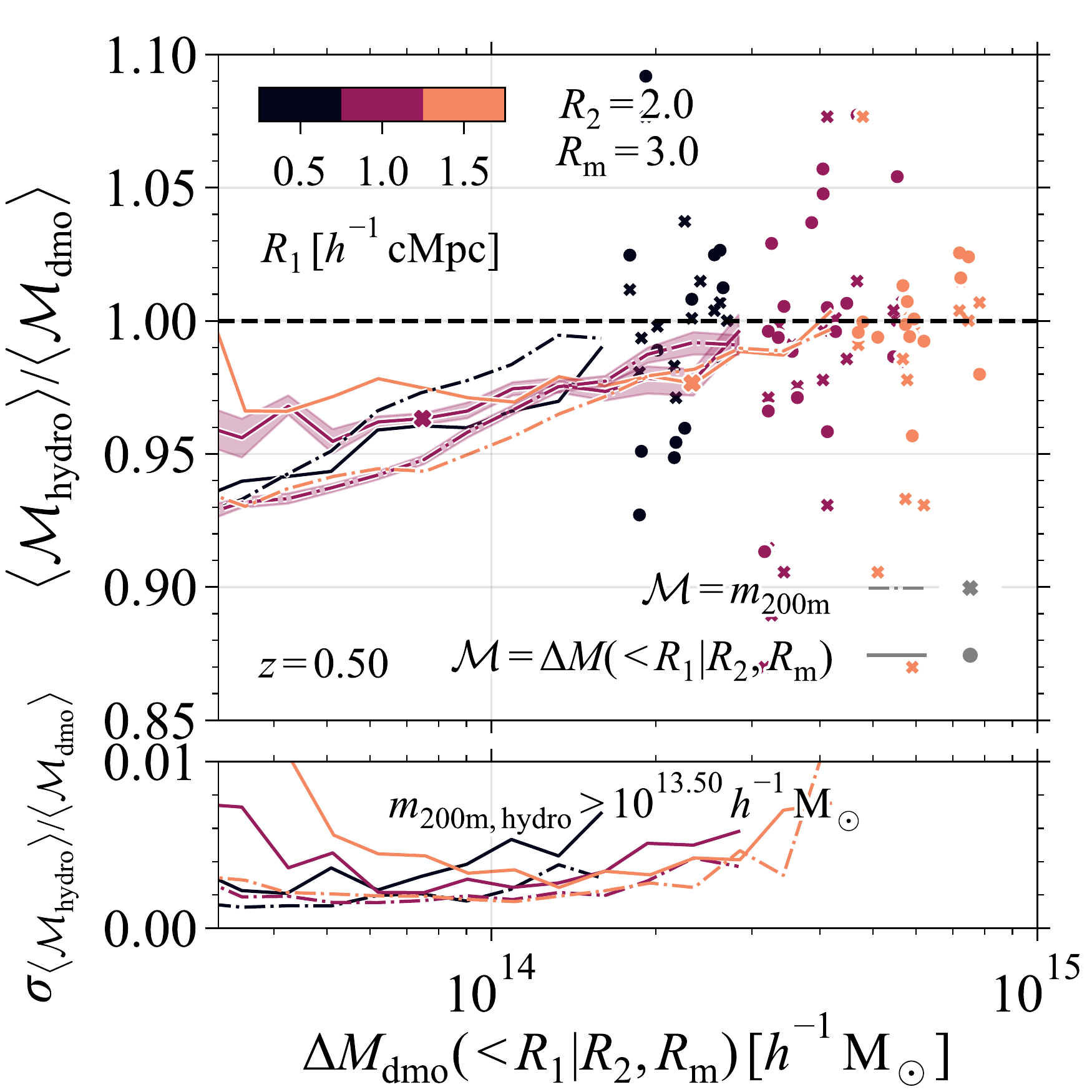}
  \caption{\emph{Top panel:} The mean aperture mass of haloes,
    measured within different apertures (solid, coloured lines), in
    the reference hydrodynamical simulation relative to their matched
    counterparts from the dark matter-only simulation at $z=0.5$,
    stacked in bins of the dark matter-only aperture mass for all
    haloes with $m_\mathrm{200m,hydro} > 10^{13.5} \, \mh$. The
    background contribution to the aperture mass is calculated within
    $\Rout=2 \, \cmpch$ and $\Rmax=3 \, \cmpch$. Coloured, dash-dotted
    lines show the 3D halo mass ratio,
    $\amean{m_\mathrm{200m,hydro}}/\amean{m_\mathrm{200m,dmo}}$,
    within the different aperture mass bins. Coloured crosses indicate
    the mean aperture mass for haloes with
    $r_\mathrm{200m,dmo} = \Rin$. Shaded regions show the bootstrapped
    error on the ratio of the mean masses for $m_\mathrm{200m}$ and
    $\Delta M$ for $\Rin = 1 \, \cmpch$. For bins with fewer than $10$
    haloes, individual measurements are shown with points and crosses
    for the aperture mass and the 3D halo mass, respectively. Aperture
    masses change less than (similarly to) 3D halo masses for haloes
    with $r_\mathrm{200m,dmo} \lesssim \, (\gtrsim) \Rin$.
    \emph{Bottom panel:} The $1\, \sigma$ bootstrapped uncertainty in
    the mass change for the different mass measures. The uncertainty
    increases for both low aperture masses with
    $\Rin \gg r_\mathrm{200m,dmo}$, dominated by matter outside the
    halo, and high aperture masses due to their lower numbers in
    BAHAMAS.}\label{fig:m_ap_ratio_selection_m3d}
\end{figure}
For a cluster cosmology analysis that uses the aperture mass function
to model the cosmology-dependence of the number counts, the relevant
mass correction is measured in bins of the aperture mass, not the 3D
halo mass. In Fig.~\ref{fig:m_ap_ratio_selection_m3d}, we show the
mean aperture (solid lines) and 3D (dash-dotted lines) mass relative
to the mean mass of matched haloes in the DMO simulation, binned by
the DMO aperture mass measured within different apertures (different
coloured lines). We only include haloes with
$m_\mathrm{200m,hydro} > 10 ^{13.5} \, \mh$ to ensure a clean cluster
sample. From Fig.~\ref{fig:m_ap_vs_m200m}, we can clearly see that
haloes with relatively low 3D halo masses can result in aperture
masses $\Delta M \gtrsim 10^{13.5} \, \mh$. However, few of these
haloes would actually be identified as clusters if we had applied an
observational cluster-finding algorithm instead of calculating the
aperture mass for all friends-of-friends haloes identified in the
simulation. For simplicity, we use the 3D halo mass in the
hydrodynamical simulations as the cluster selection criterion.

To interpret the baryonic correction when binning by the aperture
mass, $\Delta M_\mathrm{dmo}$, we identify the aperture mass with the
3D halo mass bin, $m_\mathrm{200m,dmo}$, whose haloes have the same
mean aperture mass,
$\langle \Delta M_\mathrm{dmo} | m_\mathrm{200m,dmo} \rangle$. Then,
we see that both the 3D halo mass and the aperture mass are similarly
reduced for haloes with $\Rin \lesssim r_\mathrm{200m,dmo}$ (equality
is indicated with coloured crosses). For lower-mass haloes
($\Rin > r_\mathrm{200m,dmo}$), the aperture mass is dominated by the
halo environment or structures aligned by chance along the
line-of-sight, not by the 3D halo mass, resulting in a smaller
reduction in the aperture mass than in the 3D halo mass.

Due to the large size of the BAHAMAS cluster sample, the bootstrapped
uncertainty in the mass change, shown in the bottom panel of
Fig.~\ref{fig:m_ap_ratio_selection_m3d}, is $\lesssim 1 \, \percent$
for all mass measurements. The sudden rise in the uncertainty towards
low aperture masses for the larger aperture sizes is caused by the
increased importance of matter outside of the halo, either in the halo
environment or chance line-of-sight alignments. To put this in
context, for aperture masses measured from weak lensing observations
in apertures $\Rin = [0.5, 1.0, 1.5] \, \cmpch$ and assuming a WMAP9
cosmology, the expected noise level due to the finite number of
background galaxies with an individual galaxy shape noise of
$\sigma_\mathrm{gal} = 0.3$ and a background number density of
$n_\mathrm{gal} = 30 \, \mathrm{arcmin}^{-2}$, corresponds to masses
of $\Delta M = [1.4, 3.1, 5.2] \times 10^{13} \, \mh$ (see Eq. A10 of
\citetalias{debackere2022b}), similar to the masses where the
uncertainty increases. Hence, the aperture mass range with the
increased uncertainty would most likely not be included in a
cosmological analysis.

\subsection{Scatter in the baryonic correction}\label{sec:mass_correction_uncertainty}
\begin{figure}
  \centering  \includegraphics[width=\columnwidth]{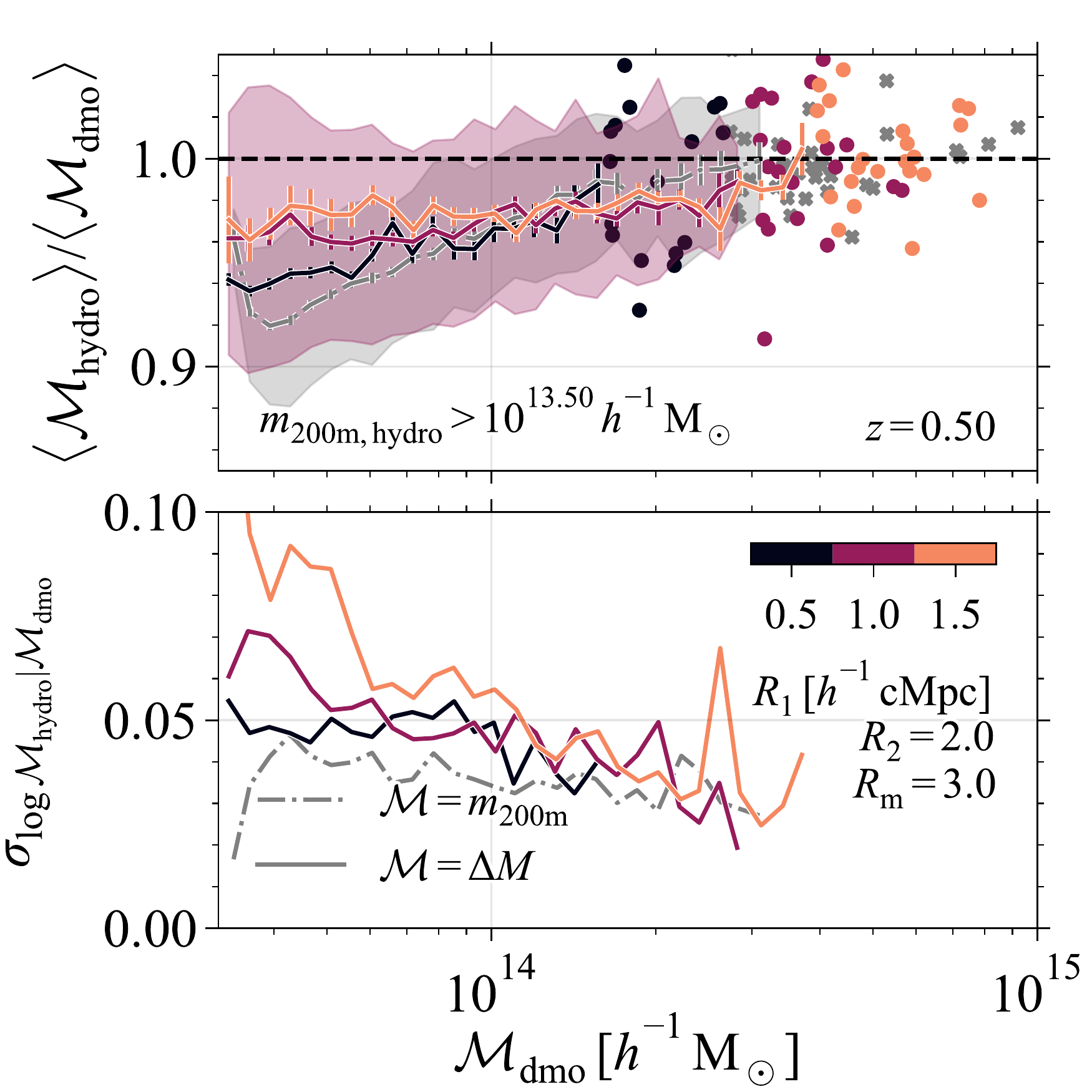}
  \caption{\emph{Top panel:} The mean bias for the aperture mass
    (solid lines) and the 3D halo mass (dash-dotted lines) as a
    function of the corresponding mass of the matching dark
    matter-only halo at $z=0.5$ for all haloes with
    $m_\mathrm{200m,hydro} > 10^{13.5} \, \mh$. Shaded regions show
    the $1 \, \sigma$ scatter for $\Rin=1 \, \cmpch$ and
    $m_\mathrm{200m}$. Error bars show the bootstrapped uncertainty on
    the mean mass bias. Coloured lines show the different aperture
    sizes $\Rin=[0.5, 1, 1.5] \, \cmpch$, with the background
    contribution calculated within $\Rout=2 \, \cmpch$ and
    $\Rmax=3 \, \cmpch$. Bins with fewer than 10 haloes show the
    individual results. Results for aperture and 3D mass measurements
    are not directly comparable since a fixed
    $\mathcal{M}_\mathrm{dmo}$ corresponds to different haloes.
    \emph{Bottom panel:} The $1\, \sigma$ scatter in
    $\mathcal{M}_\mathrm{hydro}$ at fixed $\mathcal{M}_\mathrm{dmo}$
    at $z=0.5$. For all mass measures, the scatter is smaller than
    $\approx 5 \, \percent$ for
    $\mathcal{M}_\mathrm{dmo} \gtrsim 10^{14} \, \mh$. For lower
    masses, the aperture mass scatter increases more than the 3D halo
    mass scatter due to the contribution of matter outside haloes.
  }\label{fig:m_ap_vs_m200m_uncertainty}
\end{figure}
Besides the bias in the mean mass of matched haloes in the
hydrodynamical and DMO simulations, the scatter is also important in a
cosmological analysis. If not properly accounted for, scatter in the
mass of haloes in the hydrodynamical simulation at fixed DMO halo mass
can significantly bias the cosmological parameter inference. We focus
on the 16th to 84th percentile scatter in the
$\mathcal{M}_\mathrm{hydro}$--$\mathcal{M}_\mathrm{dmo}$ relation in
Fig.~\ref{fig:m_ap_vs_m200m_uncertainty}. In the top panel, we repeat
the mean relation for both the aperture mass measurements and the 3D
halo mass, $m_\mathrm{200m}$, each binned by their respective DMO halo
mass. Hence, the aperture and 3D mass measurements cannot be compared
directly since a fixed value $\mathcal{M}_\mathrm{dmo}$ does not
include the same haloes. We show the scatter for $\Rin = 1 \, \cmpch$
and $m_\mathrm{200m}$ to compare the magnitude of the scatter to the
bias for the different mass measures.

In the bottom panel of Fig.~\ref{fig:m_ap_vs_m200m_uncertainty}, we
show the $1\, \sigma$ scatter for both the aperture mass and the 3D
halo mass, calculated as half the difference between the 84th and 16th
percentiles. For $\mathcal{M}_\mathrm{dmo} \gtrsim 10^{14} \, \mh$,
the scatter
$\sigma_{\log \mathcal{M}_\mathrm{hydro}|\mathcal{M}_\mathrm{dmo}}
\lesssim 0.05$ for all the different mass measures. Towards lower halo
masses, the scatter in the 3D halo mass stays below $0.04$, while the
scatter in the larger apertures increases to $\approx 0.07$ and
$\approx 0.1$ for $\Rin=1\,\cmpch$ and $\Rin=1.5\,\cmpch$,
respectively. The aperture mass scatter is larger due to the increased
contribution of matter outside of the halo.

In conclusion, aperture masses are, on average, slightly less
sensitive the changes in the cluster mass due to baryons because they
measure the projected density, which includes contributions from
larger radii. However, this also results in a slightly larger scatter.
A detailed comparison of a cosmology analysis using either the
aperture mass or the 3D halo mass would also need to include the
additional effect of the survey observable and its scatter at fixed
aperture mass or 3D halo mass.

\subsection{Redshift evolution}\label{sec:mass_correction_z}
\begin{figure}
  \centering
  \includegraphics[width=\columnwidth]{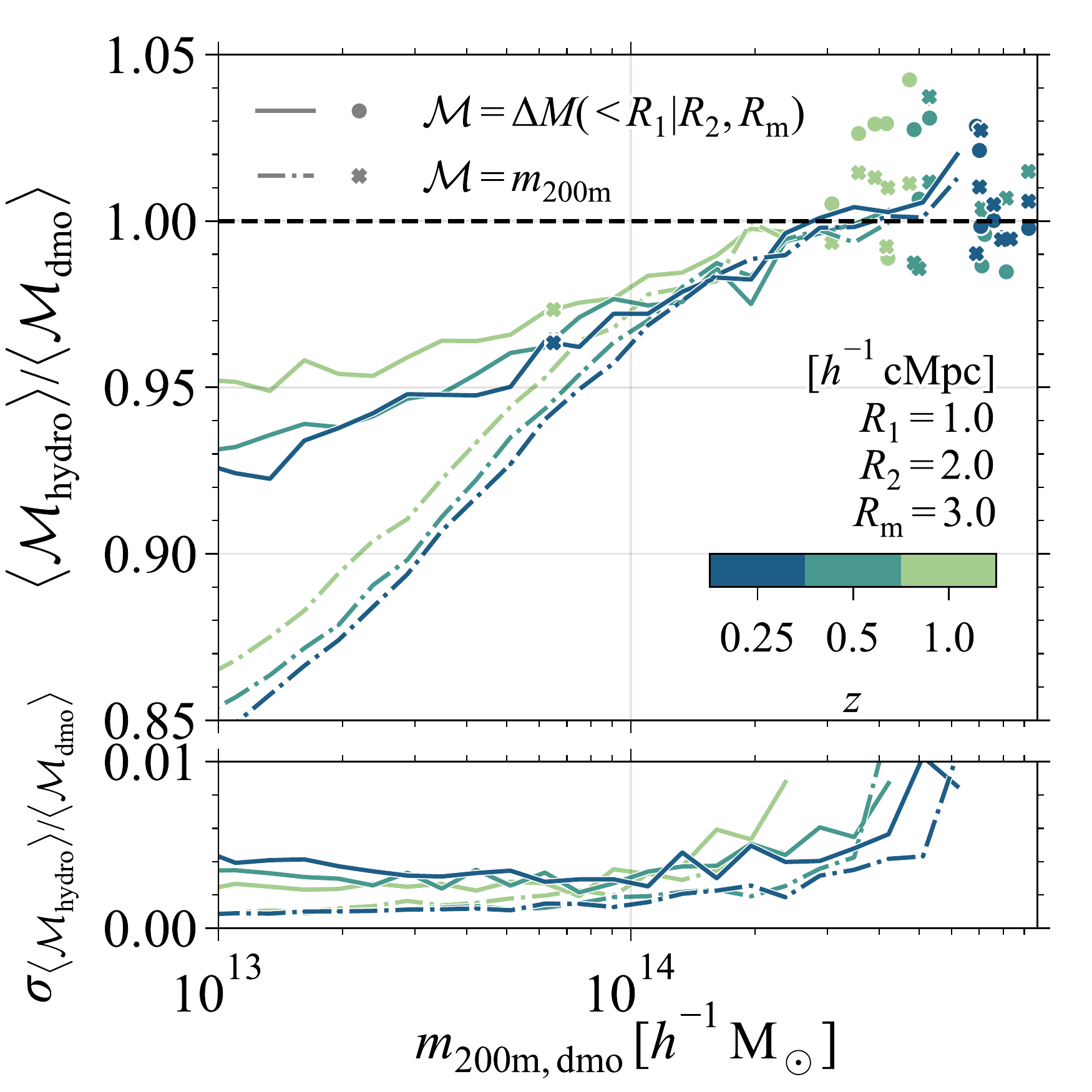}
  \caption{\emph{Top panel:} The redshift dependence of the change in
    the aperture mass,
    $\Delta M_\mathrm{hydro}(<1\cmpch|\Rout, \Rmax)$ (solid lines),
    and the 3D halo mass, $m_\mathrm{200m}$ (dash-dotted lines),
    stacked in bins of the 3D dark matter-only halo mass. All
    background correction annuli span the region between
    $\Rout=2 \, \cmpch$ and $\Rmax = 3 \, \cmpch$. Coloured lines show
    the different redshifts and crosses the halo mass for which
    $r_\mathrm{200m,dmo} = \Rin$. Bins with fewer than 5 haloes show
    the individual results. The mass reduction decreases with
    increasing redshift at fixed 3D halo mass. At all redshifts, the
    aperture mass changes less than the 3D halo mass for
    $m_\mathrm{200m,dmo} \lesssim 10^{14} \, \mh$. \emph{Bottom
      panel:} The $1\, \sigma$ bootstrapped uncertainty in the mass
    change of hydrodynamical haloes compared to their matched DMO
    counterparts for the different mass measurements. The uncertainty
    in the mass reduction only increases for the rarest, high-mass
    haloes at higher redshift.}\label{fig:m_ap_vs_m200m_zs}
\end{figure}
Since future surveys will probe clusters with high completeness and
purity up to high redshifts of $z \approx 2$ \citep[e.g.][]{adam2019},
we need to study how the impact of the inclusion of baryons on the
cluster mass changes with redshift. In
Fig.~\ref{fig:m_ap_vs_m200m_zs}, we show the redshift evolution of the
change in the aperture mass,
$\Delta M(<1 \, \cmpch| \Rout = 2 \, \cmpch, \Rmax = 3 \, \cmpch)$,
and the 3D halo mass, $m_\mathrm{200m}$, in bins of the 3D DMO halo
mass, $m_\mathrm{200m,dmo}$, between $z=0.25$ and $z=1$ where most of
the clusters will be detected. Both the aperture mass and the 3D halo
mass bias decrease slightly with increasing redshift, with the
aperture mass always being less suppressed for low-mass haloes than
the 3D halo mass. In the bottom panel of
Fig.~\ref{fig:m_ap_vs_m200m_zs}, we show that the bootstrapped
$1 \, \sigma$ uncertainty in the mean mass suppression does not evolve
appreciably.

\subsection{Dependence on feedback strength}\label{sec:mass_correction_dTheat}
\begin{figure}
  \centering  \includegraphics[width=\columnwidth]{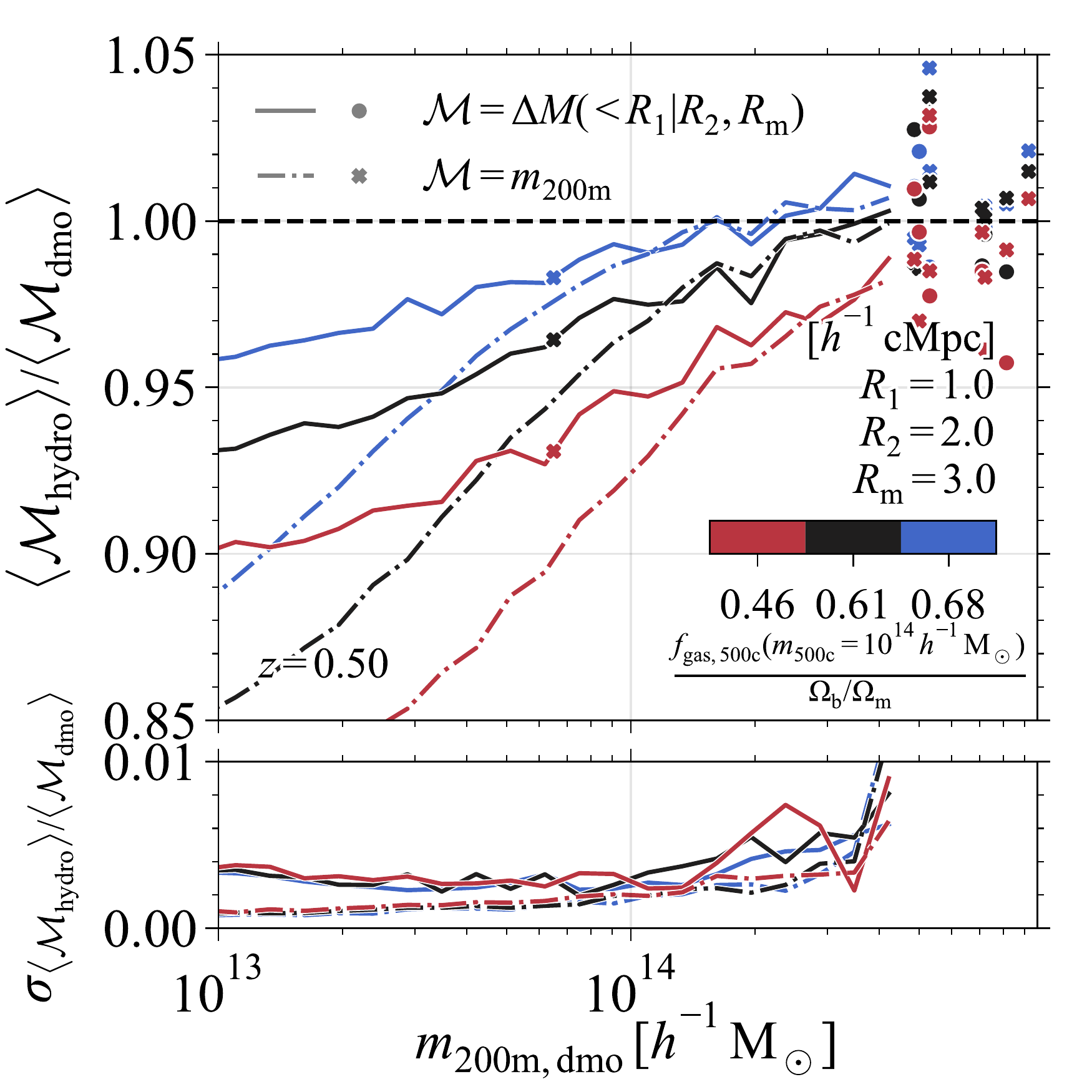}
  \caption{\emph{Top panel:} The dependence of the suppression in the
    aperture mass, $\Delta M_\mathrm{hydro}(<1\cmpch|\Rout, \Rmax)$
    (solid lines), and the 3D halo mass, $m_\mathrm{200m}$
    (dash-dotted lines) on the halo gas fraction,
    $f_\mathrm{500c,gas}(m_\mathrm{500c}=10^{14} \, \mh)$, relative to
    the cosmic baryon fraction, $\Ob / \Om$, stacked in bins of the 3D
    dark matter-only halo mass. All background correction annuli span
    the region between $\Rout=2 \, \cmpch$ and $\Rmax = 3 \, \cmpch$.
    Red (blue) lines show higher (lower) AGN heating temperatures in
    the simulation, resulting in lower (higher) gas fractions. Crosses
    indicate the halo mass for which $r_\mathrm{200m,dmo} = \Rin$.
    Bins with fewer than 5 haloes show the individual suppression
    ratios. The mass suppression increases with increasing feedback
    strength and decreasing cluster gas fractions. Aperture masses are
    consistently suppressed less than the 3D halo mass for
    $m_\mathrm{200m,dmo} \lesssim 10^{14} \, \mh$. \emph{Bottom
      panel:} The $1\, \sigma$ bootstrapped uncertainty in the mass
    suppression of hydrodynamical haloes compared to their matched DMO
    counterparts for the different mass measurements. The uncertainty
    in the mass suppression only changes slightly for the most massive
    haloes when changing the simulated AGN
    feedback.}\label{fig:m_ap_vs_m200m_dT}
\end{figure}
Finally, we study the impact of varying the strength of the simulated
AGN feedback on the cluster masses. Simulated black holes accrete from
their surrounding gas until the accumulated rest-mass energy reservoir
is sufficiently large to heat $n_\mathrm{heat}=20$ randomly chosen
neighbours to a minimum temperature $\Delta T_\mathrm{heat}$. The
fiducial subgrid parameter
$\Delta T_\mathrm{heat} = 10^{7.8} \, \mathrm{K}$ is varied to
$10^{7.6} \, \mathrm{K}$ and $10^{8.0} \, \mathrm{K}$, to have the
mean simulated cluster hot gas fractions cover the scatter inferred
from X-ray observations \citep{mccarthy2017a} while also reproducing
the galaxy stellar mass function. We point out that these variations
result in mean gas fractions that are significantly higher and lower
than the mean observed X-ray gas fractions.

In Fig.~\ref{fig:m_ap_vs_m200m_dT}, we show how the feedback strength
affects the simulated cluster mass for haloes binned by the 3D halo
mass, $m_\mathrm{200m,dmo}$. We label the simulation variations with
the true median cluster gas fraction relative to the cosmic baryon
fraction, $f_\mathrm{gas,500c} / (\Ob / \Om)$, in haloes of
$m_\mathrm{500c,hydro} = 10^{14} \, \mh$ instead of the subgrid
parameter, $\Delta T_\mathrm{heat}$, since the gas fraction can be
inferred observationally. We have not applied any post-processing to
the simulation data to include the effects of hydrostatic bias on the
cluster gas fractions inferred from observations. A higher (lower) AGN
heating temperature, shown as red (blue) lines, results in stronger
(weaker) feedback and lower (higher) cluster gas fractions. For
low-mass clusters with $m_\mathrm{200m,dmo} \lesssim 10^{14} \, \mh$,
the aperture mass is consistently affected less by the inclusion of
baryons than the 3D halo mass, while for higher-mass clusters, the
suppression is similar.

In conclusion, we have compared how the mass of clusters matched
between hydrodynamical and DMO simulations changes due to galaxy
formation processes. In particular, we showed that aperture masses are
consistently less sensitive to baryonic effects than 3D halo masses.
This property and the fact that aperture masses can be measured
directly in both simulations and observations, make the aperture mass
an excellent mass calibration tool for future cluster surveys.

\section{Conclusions}\label{sec:conclusions}
Future cosmological constraints from cluster surveys will be limited
by our understanding of the systematic uncertainty in the measured
cluster masses \citep[e.g.][]{kohlinger2015}. Since the current
standard analysis relies on theoretical predictions of the cluster
abundance based on dark matter-only simulations
\citep[e.g.][]{bocquet2019, descollaboration2020}, the modification of
the halo mass due to galaxy formation processes is one of the
systematic uncertainties that we need to take into account
\citep[e.g.][]{balaguera-antolinez2013, debackere2021}. We have used
the Baryons and Haloes of Massive Systems (BAHAMAS) suite of
cosmological, hydrodynamical simulations, which have been shown to
reproduce a wide range of the observed properties of massive systems,
to study how galaxy formation processes modify the aperture mass of
clusters compared to their matched haloes in a simulation that
includes only dark matter particles.

In agreement with \citet{debackere2021}, who studied the sensitivity
of the aperture mass to baryonic effects for idealized cluster density
profiles that reproduce the cluster hot gas fractions inferred from
X-ray observations, we find that aperture masses are less sensitive to
baryonic effects than the 3D halo mass when measured within apertures
larger than the halo virial radius. For haloes selected based on their
3D halo mass, aperture masses measured within annuli between
$1-3 \,\cmpch$, which is representative of weak lensing observations,
are consistently less suppressed by baryonic effects than the 3D halo
masses are ($\lesssim 5 \, \percent$ vs. $\lesssim 10 \, \percent$)
for all haloes with $m_\mathrm{200m} > 10^{13.5} \, \mh$
(Fig.~\ref{fig:m_ap_vs_m200m_correction}). Similar conclusions hold
when selecting haloes based on their aperture mass and ensuring that
only genuine clusters are included by using an additional lower limit
on the 3D halo mass (Fig.~\ref{fig:m_ap_ratio_selection_m3d}).

While for high-mass objects ($m_\mathrm{200m} \gtrsim 10^{14} \, \mh$)
the mass suppression due to baryons is similar for aperture and 3D
masses, we expect baryonic effects to pose a greater challenge for 3D
halo mass determinations. This is because a functional density profile
needs to be assumed to derive a 3D halo mass from observational data
and we expect this profile to be affected by baryons
\citep[e.g.][]{velliscig2014}. The smaller suppression of the mass of
group-sized haloes
($10^{13} \lesssim m_\mathrm{200m} / \mh \lesssim 10^{14}$) and the
fact that no density profile needs to be assumed to derive aperture
masses, will enable robust mass estimates and, consequently, stronger
constraints on galaxy formation processes in haloes at fixed mass.

Due to the sensitivity of the aperture mass to the halo environment,
we find that the scatter in the aperture mass in the hydrodynamical
simulation within fixed bins of the DMO aperture mass can be up to
$\approx 2$ times larger, depending on the aperture size, than the
typical scatter in the 3D halo mass
(Fig.~\ref{fig:m_ap_vs_m200m_uncertainty}). However, the scatter stays
below $10 \, \percent$ for all halo masses relevant for cluster
cosmology. Hence, this is by no means a limiting factor in the
cosmological analysis. The slightly reduced sensitivity to baryonic
effects in the cluster mass range, $m_\mathrm{200m} > 10^{14} \, \mh$,
combined with the significantly reduced systematic uncertainties in
the aperture mass measurement compared to the 3D halo mass inference,
and the high cosmological sensitivity of the aperture mass function
\citep{debackere2022b}, give the aperture mass a significant advantage
as it reduces the absolute bias due to mass calibration uncertainties
in cluster cosmology analyses.

We find only a small redshift evolution of $\lesssim 2$ \percent age
points in both the aperture mass and the 3D halo mass suppression
between $z=0.25$ and $1$ (Fig.~\ref{fig:m_ap_vs_m200m_zs}). Finally,
we find that for extreme variations in the simulation AGN feedback
strength that result in simulated \emph{mean} hot gas fractions
covering the \emph{scatter} inferred from X-ray observations of
individual clusters, the aperture mass is consistently up to
$2 \, \percent$age points less biased than the 3D halo mass, never
exceeding a suppression of $5 \, \percent$ for
$m_\mathrm{200m,dmo} > 10^{14} \, \mh$
(Fig.~\ref{fig:m_ap_vs_m200m_dT}).

Looking towards the future, calibrations of the halo mass difference
between hydrodynamical and DMO simulations can be bypassed when
large-volume cosmological, hydrodynamical simulations run for a large
grid of cosmological parameters become available. Such simulations can
be used to measure the abundance of clusters directly as a function of
any observable, avoiding the conversion between the theoretical
prediction calibrated on DMO simulations, and the true halo mass,
including the effects of baryons. Importantly, such simulations would
need to withstand thorough tests of the realism of their cluster
population. As long as such simulations are not available, however,
accounting for the effects of galaxy formation on the cluster mass is
a necessary step for any cluster cosmology survey limited in its
constraining power only by systematic uncertainties.

\section*{Acknowledgements}
This work is part of the research programme Athena with project number
184.034.002 and Vici grants 639.043.409 and 639.043.512, which are
financed by the Dutch Research Council (NWO).

\section*{Data availability}
The data used in this paper can be requested from the first author.



\bibliographystyle{mnras}
\bibliography{/Users/stijn/Documents/PhD/zotero_library}



\appendix

\bsp	
\label{lastpage}
\end{document}